\documentclass[sigconf]{acmart}

\AtBeginDocument{%
  \providecommand\BibTeX{{%
    \normalfont B\kern-.05em{\scshape i\kern-.025em b}\kern-.08em \TeX}}}

\setcopyright{acmcopyright}
\copyrightyear{2026}
\acmYear{2026}
\acmDOI{10.1145/3690000.3690000}

\acmConference[CIKM '26]{Proceedings of the 35th ACM International Conference on Information and Knowledge Management}{November 7--11, 2026}{Rome, Italy}
\acmBooktitle{Proceedings of the 35th ACM International Conference on Information and Knowledge Management (CIKM '26), November 7--11, 2026, Rome, Italy}

\begin{document}

\title{SAFE-Cascade: Cost-Adaptive Vision-Language Routing for Chart Question Answering}

%% COLUMN 1: Walmart Group A (Balanced Left Column)
\author{Ayush Dwivedi}
\email{Ayush.Dwivedi@walmart.com}

\author{Qixin Wang}
\email{Qixin.Wang@walmart.com}

\author{Ashvi Soni}
\email{Ashvi.Soni@walmart.com}

\author{Ruoteng Wang}
\email{Ruoteng.Wang@walmart.com}
\affiliation{%
  \institution{Walmart}
  \country{USA}
}

%% COLUMN 2: Walmart Group B (Balanced Center Column)
\author{Han Li}
\email{Han.Li@walmart.com}

\author{Animesh Mahapatra}
\email{Animesh.Mahapatra@walmart.com}

\author{Neeraj Agrawal}
\email{Neeraj.Agrawal@walmart.com}
\affiliation{%
  \institution{Walmart}
  \country{USA}
}

%% COLUMN 3: Academic Partner (Balanced Right Column)
\author{Xintao Wu}
\email{xintaowu@uark.edu}
\affiliation{%
  \institution{University of Arkansas}
  \country{USA}
}

\begin{abstract}

Vision-language models (VLMs) are powerful for chart question answering, but invoking a VLM for every query can be unnecessarily expensive when many questions are answerable from OCR text and lightweight language reasoning. We demonstrate SAFE-Cascade, an interactive system for cost-adaptive chart question answering. Given a chart image and a natural-language question, SAFE-Cascade first extracts chart text with OCR, obtains a provisional answer from a text-only language model, and then uses a learned router to decide whether to accept the text answer or escalate to a VLM. The demo exposes this decision process to users: OCR evidence, text-only answer, routing probability, escalation decision, final answer, estimated cost, and estimated latency are shown side by side. SAFE-Cascade is designed as a transparent interface for understanding when visual grounding is actually needed. Users can upload or select charts, ask questions, inspect the evidence used by each pathway, compare text-only and VLM answers, and adjust the escalation threshold to explore the accuracy--cost frontier. The system is implemented with Azure Document Intelligence for OCR, gpt-5-mini as the text-only model, gemini-2.5-flash-image as the VLM, and a Random Forest router trained on inference-time features. On a held-out ChartQA test split of 375 examples from a 2,500-example experiment, SAFE-Cascade achieves 69.1\% unified accuracy with 73.1\% VLM invocation, compared with 67.7\% accuracy and 100\% VLM invocation for the full-VLM baseline. The observed +1.4 percentage-point difference is statistically uncertain, so we interpret SAFE-Cascade as matching full-VLM performance while reducing VLM calls by 26.9\% and estimated cost by 9.3\%. The demonstration shows how selective modality routing can make multimodal knowledge systems more transparent, tunable, and cost-aware.

% Vision-language models (VLMs) are powerful for chart question answering, but invoking a VLM for every query can be unnecessarily expensive when many questions are answerable from OCR text and lightweight language reasoning. We present SAFE-Cascade, an interactive system for cost-adaptive chart question answering. Given a chart image and a natural-language question, SAFE-Cascade first extracts chart text with OCR, obtains a provisional answer from a text-only language model, and then uses a learned router to decide whether to accept the text answer or escalate to a VLM. The system exposes this decision process to users: OCR evidence, text-only answer, routing probability, escalation decision, final answer, estimated cost, and estimated latency are shown side by side. Users can upload or select charts, ask questions, inspect the evidence used by each pathway, compare text-only and VLM answers, and adjust the escalation threshold to explore the accuracy--cost frontier. On a held-out ChartQA test split of 375 examples from a 2,500-example experiment, SAFE-Cascade matches full-VLM performance within statistical uncertainty while reducing VLM calls by 26.9\% and estimated cost by 9.3\%. The demonstration shows how selective modality routing can make multimodal knowledge systems more transparent, tunable, and cost-aware.
\end{abstract}

%% ACM Computing Classification System (CCS) Concepts



\maketitle
\noindent{\textbf{Demo video:}} \url{https://youtu.be/ak55IiZ6rHk} \\
% \noindent{\textbf{Demo system / code / supplement:}} \url{https://example.com/project-url}

\section{Introduction}
Charts are central to knowledge communication in reports, dashboards, scientific papers, financial documents, and enterprise decision workflows. Answering questions over charts requires reading labels, extracting values, comparing visual marks, and sometimes performing numeric reasoning. Modern VLMs can solve many chart questions directly from images, but using a VLM for every query is not always necessary. Some questions can be answered from OCR-extracted text, while others require true visual grounding.

%% FIGURE 1 SHIFTED DOWN PAST THE INTRO FOOTPRINT TO PREVENT IN-PARAGRAPH FLOATING ERRORS
\begin{figure*}[t]
\centering
\includegraphics[width=0.92\textwidth]{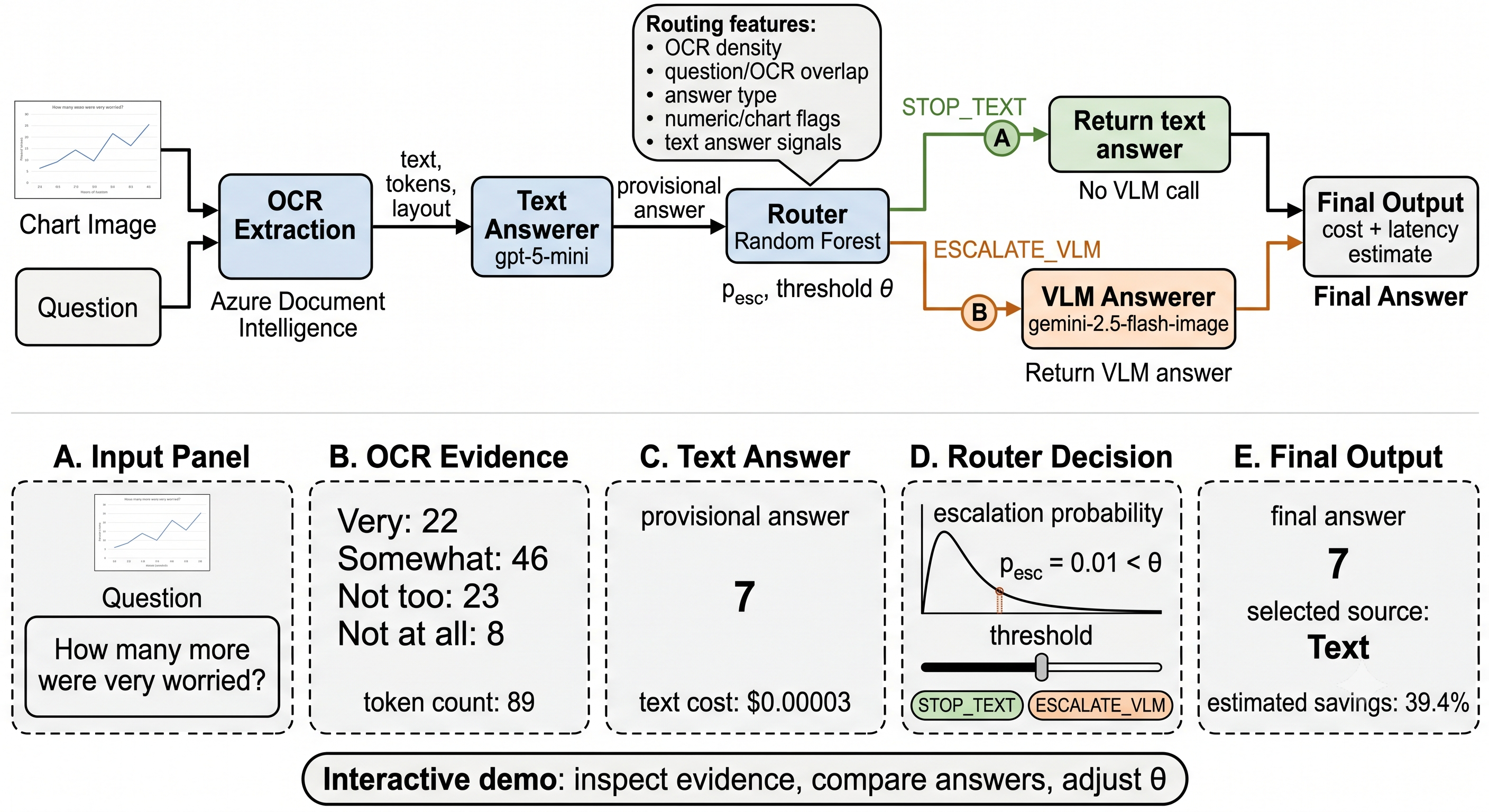}
\caption{SAFE-Cascade system architecture and processing workflow. The system exposes the same stages to users: chart/question input, OCR evidence, text-only answer, router decision, and final answer with estimated cost/latency.}
\label{fig:system}
\end{figure*}

SAFE-Cascade demonstrates a selective alternative. Instead of treating full VLM inference as the default, the system asks: does this query actually need visual inference? The system first tries a low-cost OCR-plus-text pathway. A learned router then decides whether to stop with the text answer or escalate to a VLM. The key novelty is not merely a better chart QA model; it is an inspectable modality-routing system that makes the cost, evidence, and escalation decision visible to users.

This demo targets three audiences: researchers studying cost-aware inference, document AI practitioners building OCR\slash VLM systems, and application developers deploying multimodal question answering under budget or throughput constraints. During the demo, attendees interact with chart examples, observe routing decisions, adjust the threshold, and see how cost, VLM usage, and missed-escalation risk change.

Our demonstration makes four contributions:
\begin{itemize}
    \item \textbf{Cost-adaptive chart QA system.} SAFE-Cascade routes between OCR-plus-text reasoning and VLM reasoning.
    \item \textbf{Transparent routing interface.} Users see OCR evidence, provisional answers, routing probability, final answer, and estimated cost\slash latency.
    \item \textbf{Interactive operating frontier.} A threshold slider lets users explore accuracy--cost trade-offs in real time.
    \item \textbf{Held-out ChartQA validation.} On ChartQA-2500, SAFE-Cascade matches full-VLM accuracy within statistical uncertainty while reducing VLM invocation.
\end{itemize}

\section{Positioning and Related Work}
ChartQA evaluates visual and logical reasoning over charts~\cite{masry2022chartqa}, requiring models to combine OCR, visual perception, and numeric reasoning. Document AI systems such as LayoutLMv3 and OCR-free models such as Donut improve document understanding through stronger multimodal representations \cite{huang2022layoutlmv3,kim2022donut}. SAFE-Cascade addresses a different systems question: when should a multimodal system invoke visual inference at all?

The system is also related to selective prediction, where a model abstains under uncertainty \cite{geifman2019selectivenet}, and to cost-aware LLM routing systems such as FrugalGPT and RouteLLM \cite{chen2023frugalgpt,ong2024routellm}. SAFE-Cascade differs by performing modality routing rather than only model routing. It chooses whether to invoke a vision-language pathway and exposes that decision to the user through an interactive interface.

\section{System Design}
SAFE-Cascade is a post-text routing system. Figure~\ref{fig:system} summarizes the processing workflow and the corresponding demo panels. Given a chart image and a question, the system executes four stages:
\begin{enumerate}
    \item \textbf{OCR extraction.} Azure Document Intelligence extracts chart text, layout snippets, tokens, and confidence metadata.
    \item \textbf{Text-only answering.} A text LLM receives the question and OCR text and produces a provisional answer.
    \item \textbf{Learned routing. } A Random Forest router estimates whether the text answer is reliable or whether VLM escalation is needed.
    \item \textbf{Optional VLM escalation.} If escalated, the VLM receives the chart image and question; otherwise, the text answer becomes final.
\end{enumerate}

The router uses only inference-time features: OCR length and density, question length, numeric\slash chart\slash comparison flags, question-OCR overlap, text-answer length, unknown\slash empty-answer indicators, numeric answer indicators, whether the answer appears in OCR, and estimated text-model latency. It outputs an escalation probability $p_{\mathrm{esc}}$. If $p_{\mathrm{esc}} \geq \theta$, the system escalates to the VLM; otherwise, it accepts the text-only answer. The main configuration uses $\theta=0.3$, selected on the validation split.

\subsection{Routing Objective}
SAFE-Cascade can be viewed as a cost-aware routing policy over two inference paths. For each chart-question pair, the system chooses either ($a=\texttt{STOP\_TEXT}$), returning the OCR-plus-text answer, or ($a=\texttt{ESCALATE\_VLM}$), invoking the vision-language model. The practical objective is to preserve answer quality while minimizing unnecessary VLM calls. In deployment terms, the router should stop when the text answer is likely sufficient, escalate when visual grounding is likely to improve correctness, and avoid escalation for cases where the VLM is unlikely to help. This differs from conventional selective prediction: SAFE-Cascade does not merely abstain, but acquires a stronger modality when the cheap pathway appears unreliable. The threshold ($\theta$) therefore acts as an operating knob that allows users to trade accuracy against VLM usage, cost, and latency.

\subsection{Demo Interface}
The interface is designed to make routing inspectable rather than hidden. As shown in Figure~\ref{fig:system}, it includes five panels:
\begin{itemize}
    \item \textbf{Input panel:} upload\slash select a chart and enter a question.
    \item \textbf{OCR evidence panel:} display extracted text and relevant OCR snippets.
    \item \textbf{Text-answer panel:} show the provisional OCR-plus-text answer.
    \item \textbf{Router panel:} show escalation probability, threshold, decision, and feature summary.
    \item \textbf{Output panel:} show final answer, optional VLM answer, estimated cost, and estimated latency.
\end{itemize}
The main interactive control is a threshold slider. Lower thresholds call the VLM more often and reduce missed escalations; higher thresholds reduce VLM calls but risk accepting incorrect text answers. This lets attendees see the cost--accuracy frontier directly.

\subsection{Implementation}
SAFE-Cascade is implemented as a modular service. OCR outputs are cached. Text-only answers are generated with gpt-5-mini. VLM answers are generated with gemini-2.5-flash-image. The router is a Random Forest classifier trained on split-controlled ChartQA examples. The system logs every intermediate artifact: OCR output, text prediction, router probability, VLM prediction, final answer, estimated cost, and estimated latency.

The demo is implemented as a lightweight Streamlit web application with cached OCR, text-model, VLM, and router outputs for reproducible conference presentation. The implementation emphasizes transparency and reproducibility. Latency values reported below are estimates from typical API response profiles, not measured end-to-end production timings.

\section{Evaluation}
We evaluate SAFE-Cascade on 2,500 examples sampled from the ChartQA benchmark~\cite{masry2022chartqa}, a chart question-answering dataset designed to test visual and logical reasoning over charts.\footnote{\url{https://github.com/vis-nlp/ChartQA}} We use a 70\slash 15\slash 15 split: 1,750 examples for router training, 375 examples for validation based threshold selection, and 375 held out examples for final evaluation. OCR coverage and model runs are computed over all 2,500 examples. All final accuracy, routing, cost, and error metrics are reported on the held-out test split. We use unified accuracy: an answer is correct if it satisfies normalized exact match, numeric tolerance, or ANLS similarity. Costs are estimated from configured model-call pricing.

\subsection{Main Results}
Table~\ref{tab:main} reports held-out test performance. SAFE-Cascade achieves 69.1\% accuracy, compared with 67.7\% for full VLM. Because the confidence intervals overlap, we do not claim a statistically significant improvement over full VLM. Instead, we interpret SAFE-Cascade as matching full-VLM performance while reducing VLM calls by 26.9\% and estimated cost by 9.3\%. The oracle is non-deployable and is included only as an upper bound.

Beyond matching the full-VLM baseline, the results show that learned routing is essential. The text-only path reaches only 25.3\% accuracy, and a strong heuristic cascade improves to 37.3\%, while SAFE-Cascade reaches 69.1\% on the same held-out test split. Thus, the main gain does not come merely from adding OCR or using simple visual-trigger rules; it comes from learning when the text pathway is likely insufficient. The remaining 2.6 point gap to the non-deployable oracle indicates that routing is not saturated, but the deployable router already captures most of the attainable benefit while reducing VLM calls.

%% FIGURE 2 DECLARED AS A WIDE 2-COLUMN FEATURE (figure*) FLOATING NATURALLY UNDER SECTION 4
\begin{figure*}[htbp]
\centering
\includegraphics[width=0.82\textwidth]{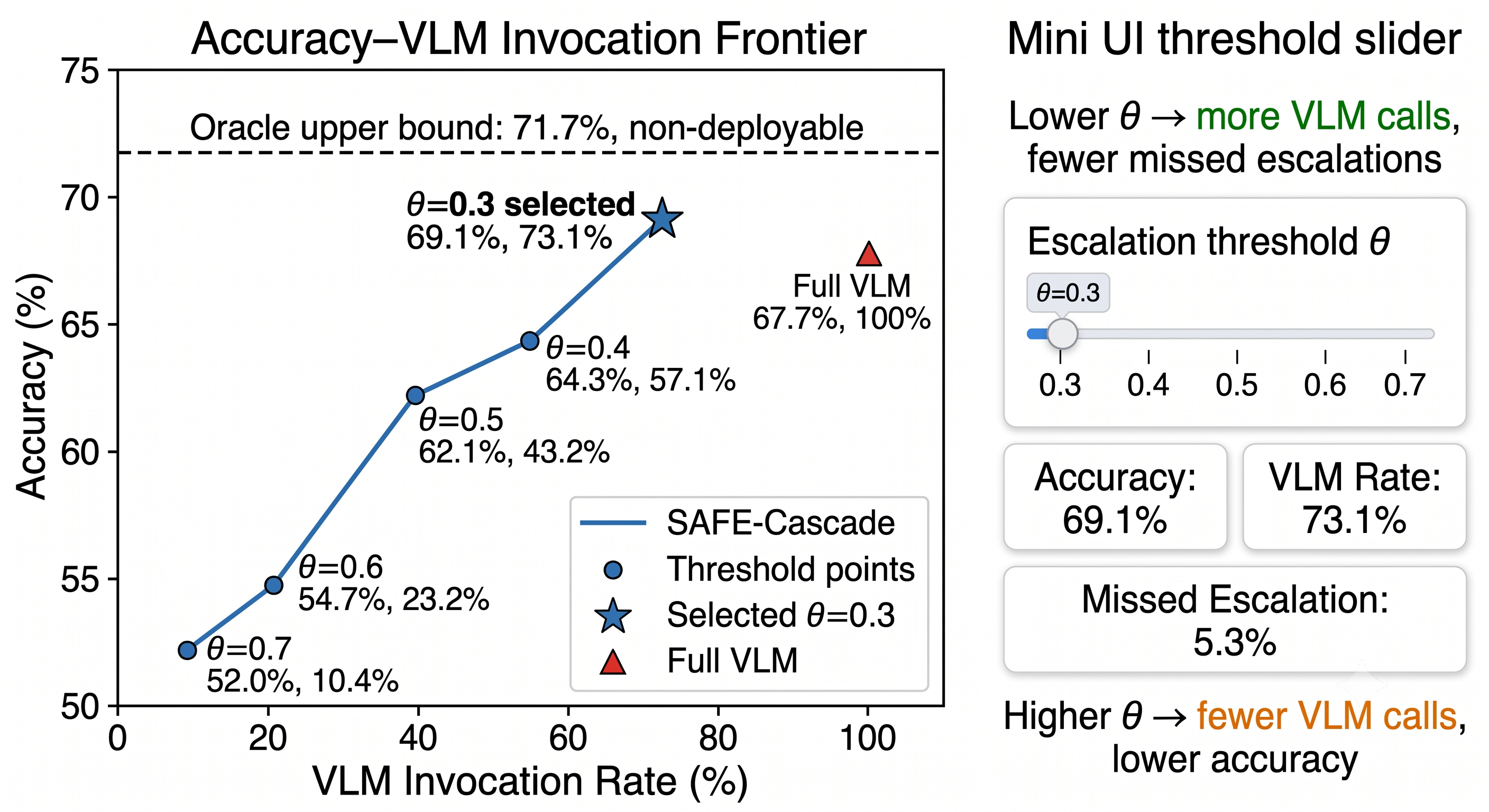}
\caption{Interactive operating frontier. SAFE-Cascade exposes the escalation threshold $\theta$ as a demo control. Lower thresholds preserve accuracy by invoking the VLM more often; higher thresholds reduce VLM usage but increase missed escalations. The selected configuration $\theta=0.3$ achieves 69.1\% accuracy with 73.1\% VLM invocation.}
\label{fig:frontier}
\end{figure*}

\begin{table}[h]
\centering
\small
\setlength{\tabcolsep}{5.5pt} 
\caption{Main results on held-out test split ($N=375$).}
\label{tab:main}
\begin{tabular}{lcccc}
\toprule
\textbf{Method} & \textbf{Acc. (\%)} & \textbf{95\% CI} & \textbf{VLM (\%)} & \textbf{Cost / 1k} \\
\midrule
Text-only & 25.3 & [20.8, 29.9] & 0.0 & \$1.23 \\
Heuristic cascade & 37.3 & [32.3, 42.4] & 30.1 & \$1.47 \\
\textbf{SAFE-Cascade ($\theta=0.3$)} & \textbf{69.1} & \textbf{[64.0, 74.1]} & \textbf{73.1} & \textbf{\$1.81} \\
Full VLM & 67.7 & [62.7, 72.3] & 100.0 & \$2.00 \\
Oracle upper bound & 71.7 & [66.9, 76.3] & 46.4 & N/A \\
\bottomrule
\end{tabular}
\end{table}

\subsection{Routing Quality}
Table~\ref{tab:routing} summarizes the router’s action-level behavior. The router achieves 81.3\% escalation precision, 87.4\% escalation recall, and 84.2\% escalation F1. The most important failure mode is missed escalation, where the system accepts an incorrect text answer even though the VLM would have been correct.

\begin{table}[h]
\centering
\small
\setlength{\tabcolsep}{8pt}
\caption{Routing behavior on held-out test split ($N=375$).}
\label{tab:routing}
\begin{tabular}{lrr}
\toprule
\textbf{Routing outcome} & \textbf{Count} & \textbf{Pct. (\%)} \\
\midrule
Correctly stopped & 60 & 16.0 \\
Correctly escalated & 152 & 40.5 \\
Missed escalation & 22 & 5.9 \\
Unnecessary escalation & 35 & 9.3 \\
Hard cases & 106 & 28.3 \\
\bottomrule
\end{tabular}
\end{table}

\subsection{Threshold Operating Frontier}
The threshold sweep is the key practical insight of the demo. As shown in Table~\ref{tab:threshold} and Figure~\ref{fig:frontier}, lower thresholds preserve accuracy by calling the VLM more often; higher thresholds reduce VLM usage but quickly increase missed escalations. The live threshold slider turns this table into an interactive experience.

\begin{table}[h]
\centering
\small
\setlength{\tabcolsep}{5pt}
\caption{Escalation threshold trade-off parameters.}
\label{tab:threshold}
\begin{tabular}{ccccc}
\toprule
\textbf{$\theta$} & \textbf{Acc. (\%)} & \textbf{95\% CI} & \textbf{VLM (\%)} & \textbf{Missed Esc. (\%)} \\
\midrule
\textbf{0.3} & \textbf{69.1} & \textbf{[64.0, 74.1]} & \textbf{73.1} & \textbf{5.3} \\
0.4 & 64.3 & [59.2, 69.6] & 57.1 & 10.7 \\
0.5 & 62.1 & [57.1, 67.7] & 43.2 & 14.9 \\
0.6 & 54.7 & [49.3, 59.7] & 23.2 & 22.9 \\
0.7 & 52.0 & [46.7, 57.1] & 10.4 & 27.2 \\
\bottomrule
\end{tabular}
\end{table}

\section{Demo Experience and Practical Insights}
SAFE-Cascade is intended to demonstrate more than a backend optimization. The interface makes multimodal inference observable. Attendees can see when OCR-plus-text reasoning is enough, when VLM escalation is necessary, why a query was escalated, how threshold changes affect VLM calls, and how cost and accuracy trade off.

The 3-minute video and live demo follow three scenarios. First, in a text-solvable case, the answer appears in OCR text; SAFE-Cascade returns the text-only answer and avoids a VLM call. Second, in a VLM-needed case, the text answer is incomplete or unreliable; the router assigns a high escalation probability and calls the VLM. Third, in a borderline case, attendees adjust the threshold slider and observe how the decision changes. For reliable review, the demo runs from cached OCR, model, VLM, and router outputs while preserving the same evidence and routing decisions as a live run.

This scenario-driven design makes the demo relevant to CIKM’s intersection of information retrieval, knowledge management, and applied AI systems. In practical deployments, the routing decision is not just an implementation detail: it determines cost, latency, user trust, and resource allocation.

\section{Conclusion and Future Work}
We presented SAFE-Cascade, an interactive system for cost-adaptive vision-language chart question answering. SAFE-Cascade treats visual context dynamically, routes between OCR-plus-text reasoning and VLM reasoning, exposes the routing decision to users, and provides a threshold-controlled interface for exploring the cost--accuracy frontier. On a held-out ChartQA test split, it matches full-VLM performance within statistical uncertainty while reducing VLM calls by 26.9\% and estimated cost by 9.3\%.

Several limitations remain. SAFE-Cascade is currently validated only on ChartQA; cross-domain validation on DocVQA, InfographicVQA, and enterprise document datasets is future work. Latency values are estimates rather than measured end-to-end timings. The current router is post-text and sequential, so future work should explore pre-routing and speculative parallel execution. Costs are estimated from configured pricing rather than production billing. The oracle is non-deployable and serves only as an upper bound. Numeric answer normalization may affect some chart QA examples.

Overall, SAFE-Cascade shows how selective modality routing can make multimodal knowledge systems more transparent, tunable, and cost-aware.

\clearpage

\section*{GenAI Usage Disclosure}
The authors used generative AI tools to assist with language editing, summarization of experimental notes, and drafting support. All system design, implementation, experimental evaluation, interpretation, and final claims were reviewed and verified by the authors.

\end{document}